# The density number of filaments in the state of the weak and optical turbulence

Laboratory of Nonlinear Optical Interactions

V.E. Zuev Institute of Atmospheric Optics Russian Academy of Sciences, Siberian Branch

A.D. Bulygin

**Annotation**
We consider the statistics of density number of filaments for the propagation of a laser beam subjected to multiple filamentation in a closed area with reflecting boundaries. Dissipation arrests the catastrophic collapse of filaments, causing their disintegration into almost linear waves.These waves form a nearly gaussian random field that seeds new filaments. The evolution of the energy distribution function of the angular spectrum and accordingly law of the dynamics of formation of the thermodynamic characteristics of the light field such as the effective temperature and entropy was found. It is established that the growth rate of the thermodynamic functions and the Hamiltonian of the system grows in proportion to the number density of filaments.
Also found that depending on the level of the average (background) intensity of the light field can move in two steady state. The first mode is realized for the steady state level of the background intensity does not exceed the value, and is characterized by typical classical (linear) of a closed thermodynamic system, a constant level of entropy and temperature, while the number density of the filaments is equal to zero. In a strongly nonlinear regime, steady state when the number density of filaments goes to a constant level, depending on a cubic, but the entropy and the effective temperature rise in a linear fashion from the distance of propagation.
The density of filaments along the propagation distance goes to a steady state, with completely destroyed the initial spatial structure of the light field. Patterns of development of the number density of filaments with access to a steady state, are nonmonotonic and significantly determined by the initial angular spectral energy distribution of the light field. The average value of filament density depends on the intensity of the complex way the linear growth stage is replaced with the output of the nonlinear saturation, nonlinear growth, corresponding to the transition to the regime of developed turbulence with growth rate in the third degree.

Propagation of high-power femtosecond laser radiation in air is accompanied by multiple filamentation, which has direct relevance to a number of important applications [1,2]. In addition, this phenomenon is one of the most important examples of the fundamental properties are manifested in stochastic nonlinear media such as plasma, solid state physics and cosmology. Despite the obvious importance of the development of statistical methods for the study of these phenomena, due to complexity of their description at the moment we can only speak about the establishment of a series of mostly qualitative patterns propagation of powerful femtosecond laser pulses. Here are some of them from the field of studying the propagation of powerful femtosecond laser radiation in the atmosphere. In the article [3] was considered the self-focusing in a Kerr media, there have been formulated for the distribution of the ratio of the intensity of the light field amplitude, and in particular, found the distribution of the number density of filaments in the form Poisson distribution.

These results seem to be considered valid only in the early stages of multiple filamentation, and low density of filaments, as discussed in this paper the model was not considered the formation of the spectral composition of the field in the decay of the filament and thus the formation of filaments subsidiaries of the parent [4] . In [5] was established form of the energy distribution of the amplitudes in the steady state for an open system with amplification. This feature of the distribution may thus be regarded as a characteristic feature of the optical turbulence regime.

Formulation of the problem considered in this work is largely similar to the problem discussed in the papers [5], but there are significant differences. First, the nonlinear mechanism for stopping the collapse of a stationary model was chosen from the condition of compliance with the formation of the angular spectrum, or in other words, the ring structure for filamentation in the nonstationary case.

So how exactly is the structure responsible for the development of the filamentation of the next generation [4]. In contrast to [5], were selected non-linear mechanism for stopping the collapse is a steep and, accordingly, this ring structure is more pronounced bright.

Second, in our work will be considered a conservative closed energy system, which will ensure closure of reflecting boundaries, this is a classic statement of the problem in the thermodynamic sense

Such a formulation of the problem is useful, not only in terms of fundamenalnoy research, but also with the application as it allows to establish regularities in the formation of multiple filamentation at different fixed values of the intensity of the background field. Also, the establishment of such patterns is important in terms of modeling of multiple filamentation in the atmosphere.

The aim of this work will establish patterns of entry mode steady state at different values of the background intensity of the laser radiation, as well as analysis of the formation characteristics of the light field corresponding thermodynamic quantities [6] when entering this mode.

Here we describe the propagation of a laser beam through the amplified Kerr media by the regularized nonlinear Schrödinger equation (RNLS) in dimensionless form:

$$\frac{\partial}{\partial z} U(\mathbf{r}_\perp, z) = \left( \frac{i}{2n_0 k_0} \nabla_\perp^2 + ik_0 I \right) U \quad (1)$$

and

$$\varepsilon_{eff}(I/I_{cr}, \chi) = -n_2 \left( (1 + I/I_{cr}) e^{-\chi I/I_{cr}} - I/I_{cr} \right).$$

Where the beam is directed along the z axis, $\mathbf{r}_\perp \equiv (x, y)$, are the transverse coordinates, is the envelope of the electric field $U(\mathbf{r}_\perp, z)$, and $\nabla_\perp \equiv (\frac{\partial}{\partial x}, \frac{\partial}{\partial y})$; $k_0 = 2\pi/\lambda_0$ - wave number ($\lambda_0 = 800$ нм); $n_2$ - cubic nonlinearity of the refractive index of air, respectively the critical power of self-focusing [1,2] for air $P_{cr} = 2\pi/(k_0^2 n_2) \approx 3,2 Gw$); $\varepsilon_{eff}$ - is adjustable function of the effective nonlinearity, which for the values of the intensity of light $I \equiv |U|^2$ is much less than $I_{cr}$ the will becomes nonlinear Kerr-type nonlinear models for a defocusing medium, the value $I_{cr}$ chosen in accordance with the standard value of the intensity for filamentation in air $5 \cdot 10^{13} W/cm^2$, an adjustable parameter "sharpness" was chosen from the condition of best fit the profile of the angular spectrum, which is formed after the passage in nonlinear focus, in the case of the solution of the full time-dependent formulation of (NLS) [2,4], in terms of maximum intensity, it turned out that the best agreement is reached at $\chi \to \infty$.

With the goal is to provide simulation of the reflection from the boundary wall, the effect of profit used total internal reflection. To do this, on the borders of the refractive index is chosen to be zero. This method will allow a very efficient numerically implement mirrored border.

Statistical properties of the classical linear statistical views are completely characterized by the energy distribution function in velocity [6] to describe the light field, this corresponds to the distribution function of the angular spectrum [7], or the Wigner function [8], defined as:

$$\rho(k, z) = |U(\mathbf{k}, z)|^2 / \int_k d\mathbf{k} |U(\mathbf{k}, z)|^2$$

Where $U(\mathbf{k}) \equiv \int_{-\infty}^{\infty} U(r) e^{-ikr} d\mathbf{k}/2\pi$ - is the Fourier image of the light field $U$. Important characteristics of the distribution function is its second moment :

$$\theta^2(z) \equiv \int_{-\infty}^{\infty} \rho(k, z) \mathbf{k}^2 d\mathbf{k} / \int_{-\infty}^{\infty} \rho(k, z) d\mathbf{k},$$

which has in the optical sense of the angular divergence, as in classical statistical physics, this quantity corresponds to the temperature of the light field [7]. Another important feature is functional is a functional entropy:

$$s(z) \equiv \int_{-\infty}^{\infty} \rho(k, z) \ln(\rho(k, z)) d\mathbf{k} / \int_{-\infty}^{\infty} \rho(k, z) d\mathbf{k}.$$

Along with these characteristics of the light field, the nonlinear medium is important for the energy distribution in the amplitude of the light field $\rho_I(I,z)$, or equivalently the distribution of intensity. Example of input variables in the propagation of the light field in a closed system is shown in the graphs (Fig.1)

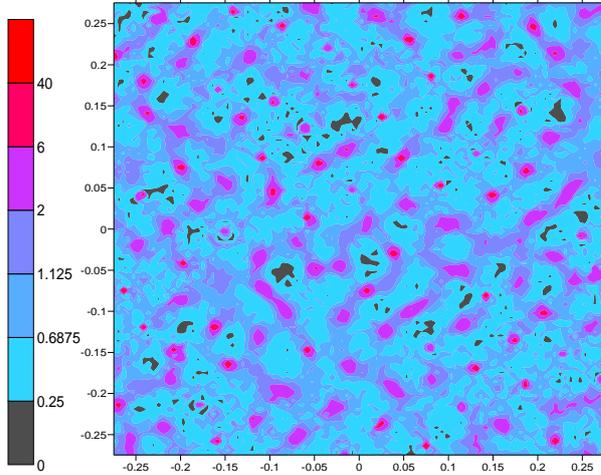

Fig. 1. An example of the intensity normalized intensity of the light field in the optical turbulence versus spatial coordinates x/R0; y/R0

On Fig. 2 shows an example of the density of filaments and the above thermodynamic parameters of the spread of distance to the background intensity $2,55 \cdot 10^{11}\ W/cm^2$ and the initial Gaussian noise of 10%.

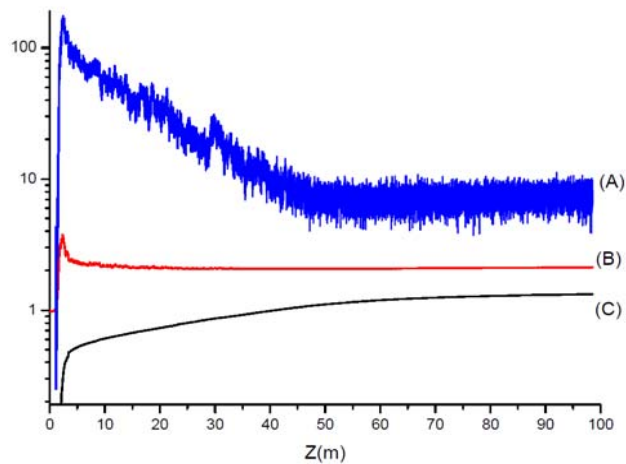

Fig 2. (A) - The dependence of the number density of filaments $n_f (cm^{-2})$; (B) - "entropy" of the light field $s$ and the square effective width of the angular spectrum of the light field "temperature» $\theta^2$ (C) of the distance distribution for the case of the intensity of the background field $2,55 \cdot 10^{11} W/cm^2$.

View of distribution of Fig. (3) in our model, in general, almost the same as that described in [5] that, in general, due to the presence of a Kerr medium, and mechanism for stopping the collapse of the value is realized when the intensity is much higher than the background level of intensity values.

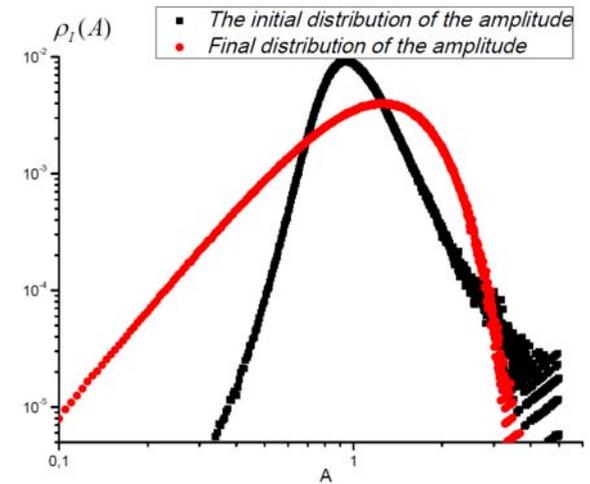

Fig. 3. The energy distribution of the light field amplitude is almost at the beginning of the propagation path (through 1.65 m), black curve, the energy distribution of the light field amplitude near the end of the propagation path (through 1.65 m) red curve (at 100 m).

Let us consider the energy distribution of the light field on the angular spectrum $\rho(k,z)$. As can be seen from the graphs of Fig. (4), the propagation of the light field is accompanied by a transformation of the angular spectrum and the allocation of the two main modes of peaks in the spectral distribution.

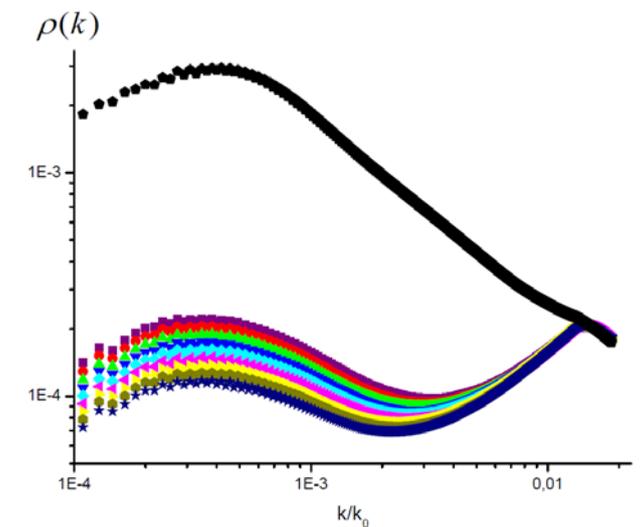

Fig 4. The energy distribution of the light field on the angular spectrum at different distances of propagation. The black curve is the distance (through 1.65 m), colored curves correspond to the shape of the spectrum in the regime of nonlinear steady state of 80 m (purpornye squares) to 100 m (blue star) with an equal interval.

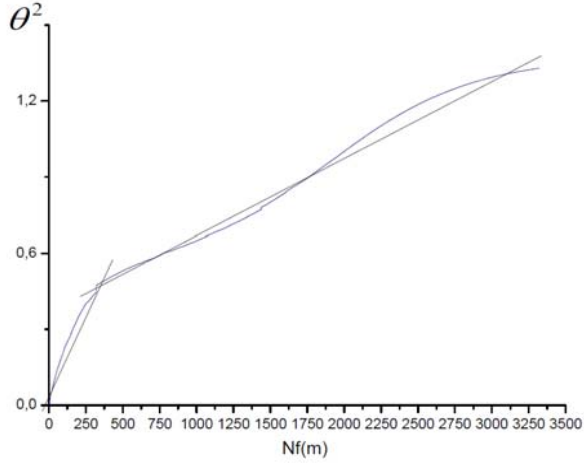

Fig. 5 The effective divergence of the light field (temperature) of the integral value of the number of filaments along the propagation distance $N_f = \int_0^z n_f(z')dz'$.

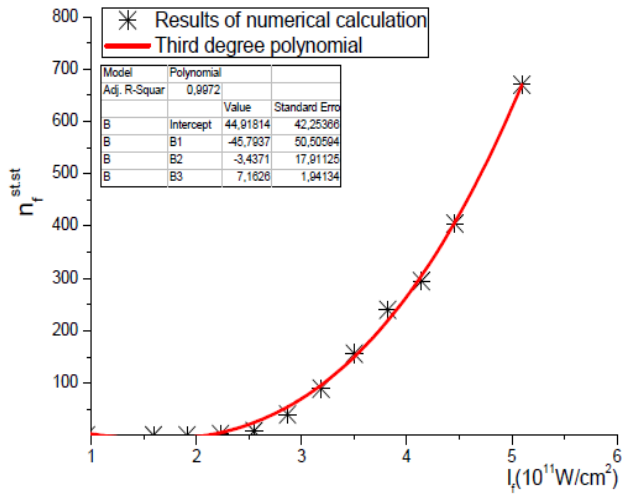

Fig. 6 The dependence of the number density of filaments $n_f^{st.st}$ in a steady-state mode, the value of the background intensity of the light field

First, relatively low-frequency peak corresponds to the scale of the transverse filament or focus in a stationary model. The second peak is already determined the choice of the model to stop the collapse, since it corresponds to a ring structure formed at filamentation so than the "harder" to stop the collapse of the model, the stronger the peak is pushed to the high-frequency part of the spectrum. It is clear that in the process filamentation is "pumping" the energy of the light field in the high-frequency part of the spectrum, as is evident from Fig. 4.

As can be seen from the Fig. 5 dependence $\theta_e^2$ on the distance spread quickly acquires a linear form. It describes the type of pattern formation in the effective divergence of the output in the nonlinear regime, see the Fig. 2, i.e. of linear growth. Behaves similarly, and "entropy."

The number filaments grows until reaching a statistical steady-state corresponding to fully developed optical turbulence [2,5,9] Fig. 2. As can be seen from these graphs Fig. 2 and Fig. 6, we can distinguish two fundamentally different modes of the stationary state, the intensity of the background level of demarcation that separates these two regimes is $I_{cr} \approx 2,5 \cdot 10^{11} W/cm^2$. The first steady state regime is characterized in many ways the classic mode of thermodynamic equilibrium in the linear case. That is, the constant values of the effective temperature and entropy and the zero-density filaments. Differences appear only in the form of energy distribution in the spatial modes in Fig. 4 because of the nonlinearity of Kerr type.

Second, the regime is characterized by a quasi-level density of the filaments, and thus a linear temperature increase in entropy. A detailed discussion of the second mode is essentially a nonlinear steady state.

As can be seen from the graph of Fig. 6 upon reaching the critical value of intensity $I_{cr}$, the growth of the number density of filaments in the equilibrium state depending on the level of background intensity, determined cubic degree. In this case the effect of entering the saturation regime the number density of filaments in a steady state mode, depending on the intensity he has not been observed. That, however, is not contrary to that found in the saturation effect, since there is no reason to believe that in the case considered in the regime of steady state.

Indeed the development of multiple filamentation in the hot regions leads to a significant increase in the diffraction divergence in these hot areas, making them " resorption" even before the filament density in the developed state of the steady state.
Of course, it is worth noting here that talk about the dependence of the number density of filaments from the background intensity is valid only for steady state.

The existence of the second regime steady state, with distance less and less a physical sense, since we are in the paraxial approximation [2,9].

Accordingly, the formal broadening of the spectrum or the growth of entropy and temperature, ceases to lose the physical meaning of the more than in the angular distribution of the spectrum of the light field contains more high frequency components.

**References**

1. *J.Kasparian, J.-P.Wolf* Physics and applications of atmospheric nonlinear optics and filamentation// Optics Express, 2008, V. 16, № 1. P.466–493.
2. *R.W. Boyd, S.G. Lukishova, Y.R. Shen* Self-focusing: Past and Present *// Topics in Applied Physics. Vol. 114., New York: Springer.,2009., p. 605.*
3. *J.Garnier*// Phys. Rev. E., **73**, 046611 (2006)
4. *Kosareva O.G., Panov N.A., Kandidov V.P.* // Atmospheric and Oceanic Optics Journal V.E. Zuev Institute of Atmospheric Optics. **18**, 204 (2005)
5. *Pavel M. Lushnikov, Natalia Vladimirova* // Opt. Lett., **35**, 1965 (2010)
6. *R. Balescu* Equilibrium and Nonequilibrium Statistical Mechanics (New York: Wiley, 1975)
7. *S. A. Akhmanov, Y. E. D'yakov, and A. S. Chirkin,* Introduction to Statistical Radiophysics and Optics *(Nauka, Moscow, 1981),* p. 306.
8. *E .Wigner*// Phys. Rev. E., **40**, 749 (1932)
9. S. Henin , Y. Petit , J. Kasparian , J.-P. Wolf et al// *Appl. Phys. B* **100**, 77–84 (2010)